\pdfoutput=1
\documentclass[%
 aip,
 sd,%
 amsmath,amssymb,
 reprint,%
]{revtex4-1}
\usepackage{graphicx}
\usepackage{dcolumn}
\usepackage{bm}
\usepackage{siunitx}

\begin{document}

\preprint{AIP/123-QED}

\title[Time-resolved measurement of single pulse femtosecond laser-induced periodic surface structure formation]{Time-resolved measurement of single pulse femtosecond laser-induced periodic surface structure formation}%

\author{K. R. P. Kafka}
\author{D. R. Austin}
\affiliation{ 
Department of Physics, The Ohio State University, 191 W. Woodruff Ave, Columbus, OH 43210, USA}

\author{H. Li}
\author{A. Yi}
\affiliation{Department of Integrated Systems Engineering, The Ohio State University, 1971 Neil Ave
Columbus, OH 43210}

\author{J. Cheng}
\affiliation{Harbin Institute of Technology, 92 Xidazhi St, Nangang, Harbin, Heilongjiang, China, 150001}

\author{E. A. Chowdhury}
\affiliation{ 
Department of Physics, The Ohio State University, 191 W. Woodruff Ave, Columbus, OH 43210, USA}

\date{\today}

\begin{abstract}
Time-resolved diffraction microscopy technique has been used to observe the formation of laser-induced periodic surface structures (LIPSS) from the interaction of a single femtosecond laser pulse (pump) with a nano-scale groove mechanically formed on a single-crystal Cu substrate.  The interaction dynamics (0-1200 ps) was captured by diffracting a time-delayed, frequency-doubled pulse from nascent LIPSS formation induced by the pump with an infinity-conjugate microscopy setup. The LIPSS ripples are observed to form sequentially outward from the groove edge, with the first one forming after 50 ps. A 1-D analytical model of electron heating and surface plasmon polariton (SPP) excitation induced by the interaction of incoming laser pulse with the groove edge qualitatively explains the time-evloution of LIPSS formation.
\end{abstract}

\maketitle

Femtosecond laser-induced periodic surface structures (LIPSS, or ripples) have been actively studied during the last decade, due to applications of surface structuring such as laser micro/nano machining \cite{Chimmalgi2005}, solar cells \cite{Sarnet2008}, wave guides \cite{Davis1996}, super hydrophobic surfaces \cite{Wu2009}, and many others \citep{Huang2007,Vorobyev2008}. 

While the mechanism by which femtosecond pulse interaction leads to LIPSS formation on different materials is under debate \cite{Reif2014}, the model for describing the period and orientation of LIPSS with spatial period $\Lambda>\lambda/2$ (laser wavelength $\lambda$) is held in strong consensus \citep{Sipe1983,Huang2009}. In this model, the incident laser excites and interferes with a surface electromagnetic wave/surface plasmon polariton (SPP), and creates periodic intensification of the fields at the interface, which somehow results in groove formation with the same period. While SPPs can only be excited natively in metals, LIPSS also may be formed on dielectrics and semiconductors given sufficient laser-induced modification of the dielectric function via inter-band transition, so as to become metallic during the laser pulse\citep{Huang2009}. In addition to metallicity, SPP excitation from free-propagating radiation requires a surface structure or roughness. Femtosecond LIPSS formation is therefore understood essentially as a multi-pulse effect, where a phase-matched grating structure is developed at $\Lambda$ only after of several pulses of ablation undergoing positive feedback \citep{Miyazaki2014}.

The timescale of LIPSS formation is fundamentally interesting for femtosecond pulses because both the laser and SPP ($\sim\SI{}{\femto\second}$)would have long since vanished before the expected timescale of material movement ($\sim\SI{}{\pico\second}$). And while dynamics of the femtosecond laser damage process (without LIPSS) has been studied using pump-probe techniques \cite{MacDonald2007}, the vast difference between the damage timescale and typical laser repetition rates ($\SI{}{\nano\second}\sim\SI{}{\milli\second}$) has limited these measurements to dynamics primarily of single-pulse damage. Consequently, the multi-pulse nature of femtosecond LIPSS makes them a challenging subject for a pump-probe study. So far there has only been one effort to temporally resolve their formation \citep{Murphy2013pumpprobe} where Murphy et. al. used a double-pump pulse and probe, to observe some evidence of periodic structure formation on Si surface at +50 ps, but they were not able to follow the dynamics afterward and the whole interaction was complicated by non-uniform damage crater and possibly residual heating/incubation created by the first pump pulse. In this work, we present a method to produce LIPSS with a single femtosecond pulse by way of a mechanically-formed groove on a copper target in air, and dynamically resolve LIPSS formation with Time Resolved Diffraction Microscopy (TRDM) for the first time. 

\begin{figure}
\includegraphics[scale=.2]{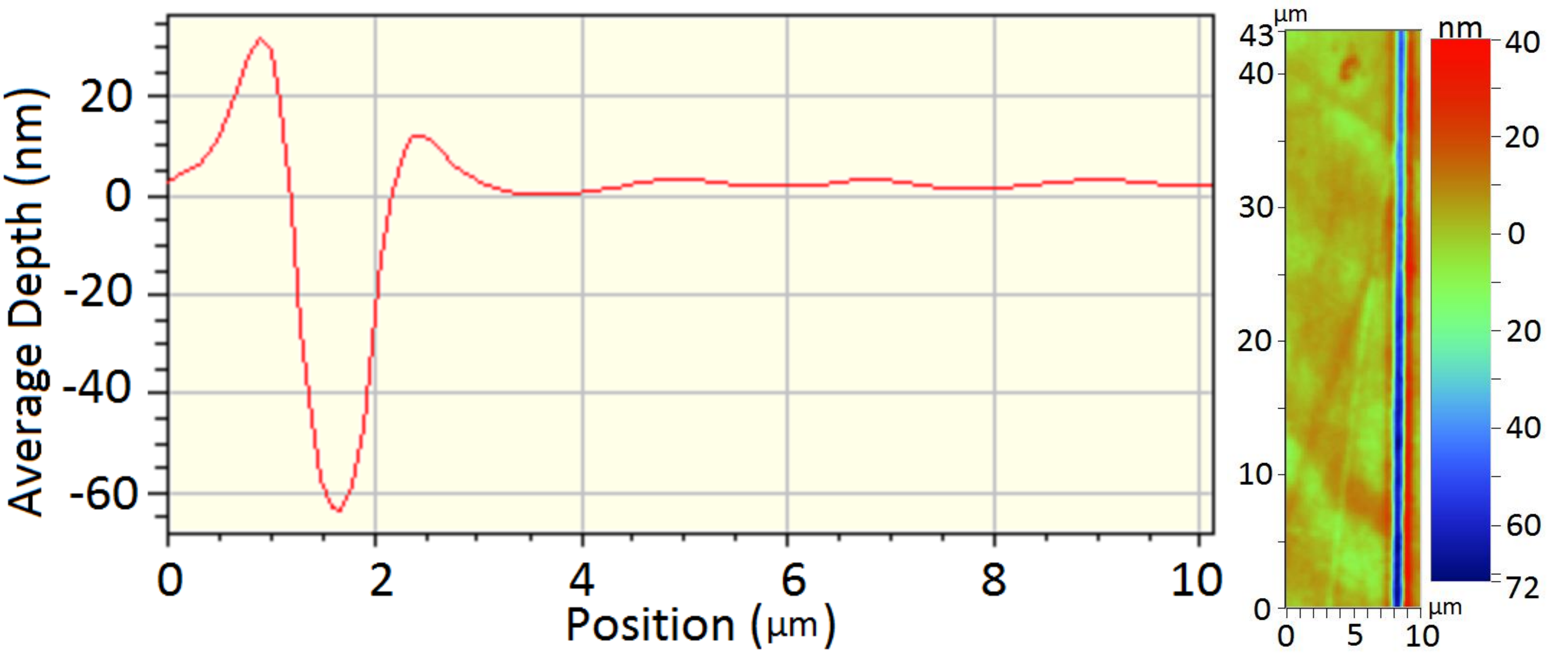}
\caption{\label{fig:groove}Depth profile (Wyko) of machined groove: $\SI{1}{\micro\meter}$ wide, $\SI{100}{\nano\meter}$ peak-to-valley. Depth is averaged over a $\SI{10}{\micro\meter}$ length.}
\end{figure}

For this experiment, single crystal Cu (MTI Corp. $10\times10\times\SI{1}{\milli\meter}, \langle100\rangle$) was chosen as a target instead of dielectric or semiconductor targets like sapphire or Si because the dielectric function of Cu naturally provides support for SPP creation and eliminates other time-dependent variables such as rapidly changing electron density due to interband transition and ionization of electrons from valence bands causing strong spatio-temporal variation in excited di-electric function in non-metals. To couple SPPs in Cu, a series of well-characterized nano-scale grooves were machined with a micro-structured diamond tool (overall nose radius $\SI{234}{\micro\meter}$)using a broaching process (tool moves in a linear fashion to create the grooves repeatedly) on the 350 FG (Freeform Generator, Moore Nanotechnology, Inc.). This produced a nearly-identical family of grooves with $\SI{100}{\micro\meter}$ pitch and average depths ranging from $\SI{100}{\micro\meter}$ to $\SI{100}{\nano\meter}$, with progressively narrower profile and shallower depth. For our experiment, the site for LIPSS formation was the outermost of each family of grooves, which was $\SI{1}{\micro\meter}$ wide and $\SI{100}{\nano\meter}$ deep, as shown in Fig. \ref{fig:groove}. The nearest neighbor groove was $\SI{8}{\micro\meter}$ to the right (not seen in fig.). To avoid any influence from the other satellite grooves, the laser focus was positioned $\sim\SI{25}{\micro\meter}$ to the left of the nano-groove, which was over 75 microns away from the nearest next series of grooves on the left. 
\begin{figure}
\includegraphics[scale=.51]{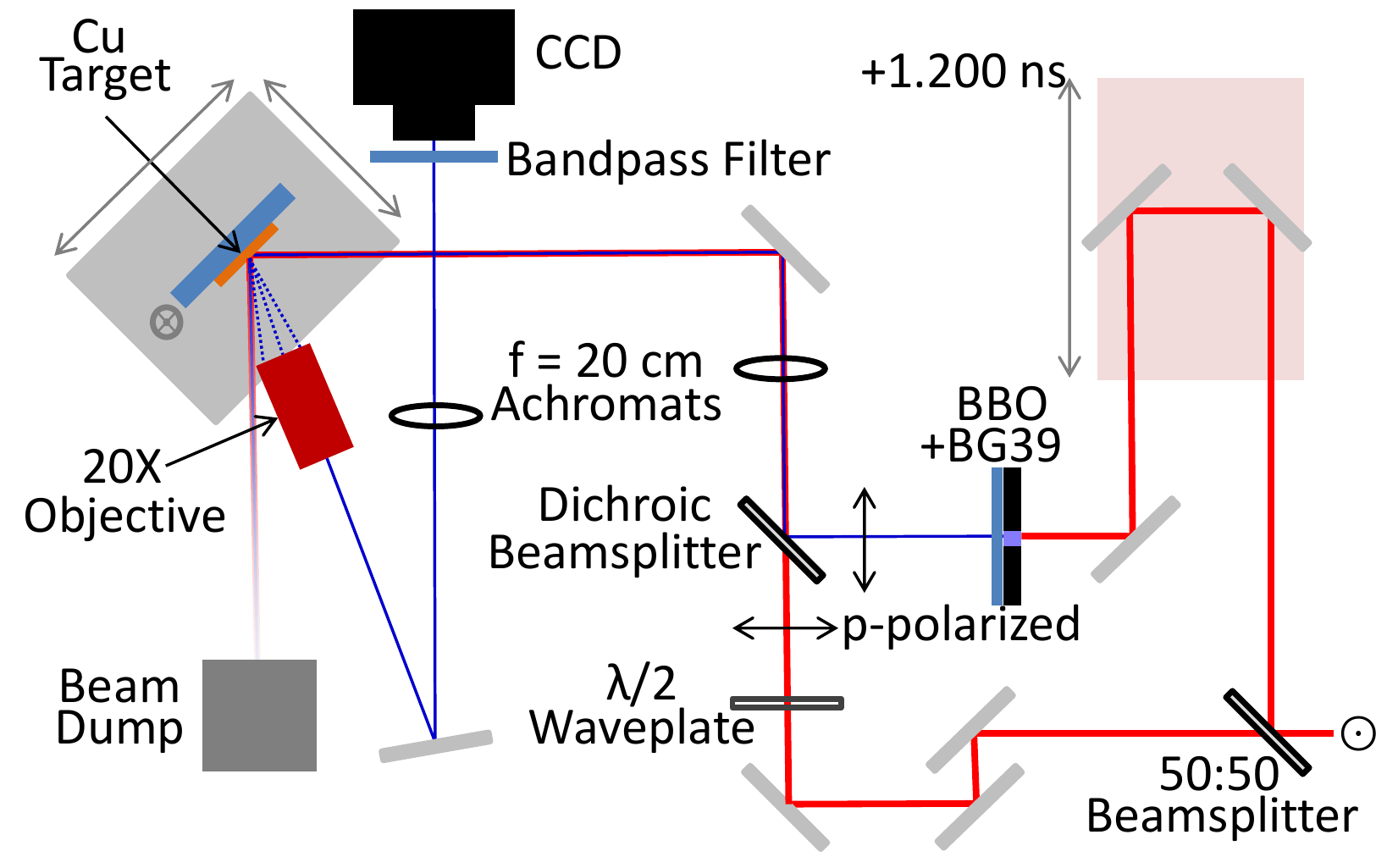}
\caption{\label{fig:expsetup}Time-resolved diffraction microscopy experimental setup. A single pulse from the pump beam at $\SI{45}{\degree}$ AOI forms LIPSS on the Cu target near a machined nano-groove. A microscope objective images LIPSS formation under illumination of the diffracted frequency-doubled probe beam at specified time delay. The objective lens is oriented at $\SI{12}{\degree}$ from normal to gather more diffracted signal. Sample is translated along groove structure to a fresh site for the next pump pulse.}
\end{figure}

A home-built Ti:sapphire liquid-nitrogen-cooled regenerative amplifier produced laser pulses at 500 Hz, with central wavelength 773 nm, and pulse energy $>\SI{3}{\milli\joule}$, which were compressed to 60 fs pulse length. Single pulses extracted from this train using an external Pockel’s cell were sent into the setup shown in Fig. \ref{fig:expsetup}. The beam was split into a pump arm and a frequency-doubled probe arm, with the probe's delay stage (Newmark Systems NLS4-8-16-E1) variable from -0.100 ns to +1.200 ns. The two beams were recombined with a dichroic beamsplitter (Semrock Di02-R405) and focused onto the target at 45 degrees angle of incidence (AOI) in $p$-polarization using an $f=\SI{200}{\milli\meter}$ achromatic lens. The pump beam focus had a Gaussian waist diameter of $2w_0=\SI{43}{\micro\meter}$ and was filtered to $4.47\pm\SI{0.06}{\micro\joule}$ per pulse, corresponding to a beam-normal peak fluence $\SI{0.61}{\joule/\centi\meter^2}$. Before inserting the target, pump-probe zero-delay timing was determined by mixing them in a $\SI{100}{\micro\meter}$ thick nonlinear crystal (BBO) to produce third-harmonic signal. The damage site is illuminated by probe light incident at a known time delay, which is observed by an in-situ 20x infinity-corrected microscope objective (Mitutoyo) and imaged onto a triggered CCD camera (Basler scA1600-14gm). Due to their periodic nature, light scattering from LIPSS tends to resemble diffraction from a grating, while the rest of the substrate almost exclusively reflects in the specular direction. Then by positioning the objective near $\SI{12}{\degree}$ from target-normal, we simultaneously rejected the 0-order reflection and collected diffractive orders 2-4, resulting in high signal-to-noise for the imaging of the LIPSS. While plasma self-emission is often a nuisance in pump-probe imaging experiments, by choosing a low enough pump fluence in combination with a bandpass filter (Semrock 390/40), the pump contamination is observed to be below the noise threshold of the CCD.
\begin{figure}
\includegraphics[scale=.28]{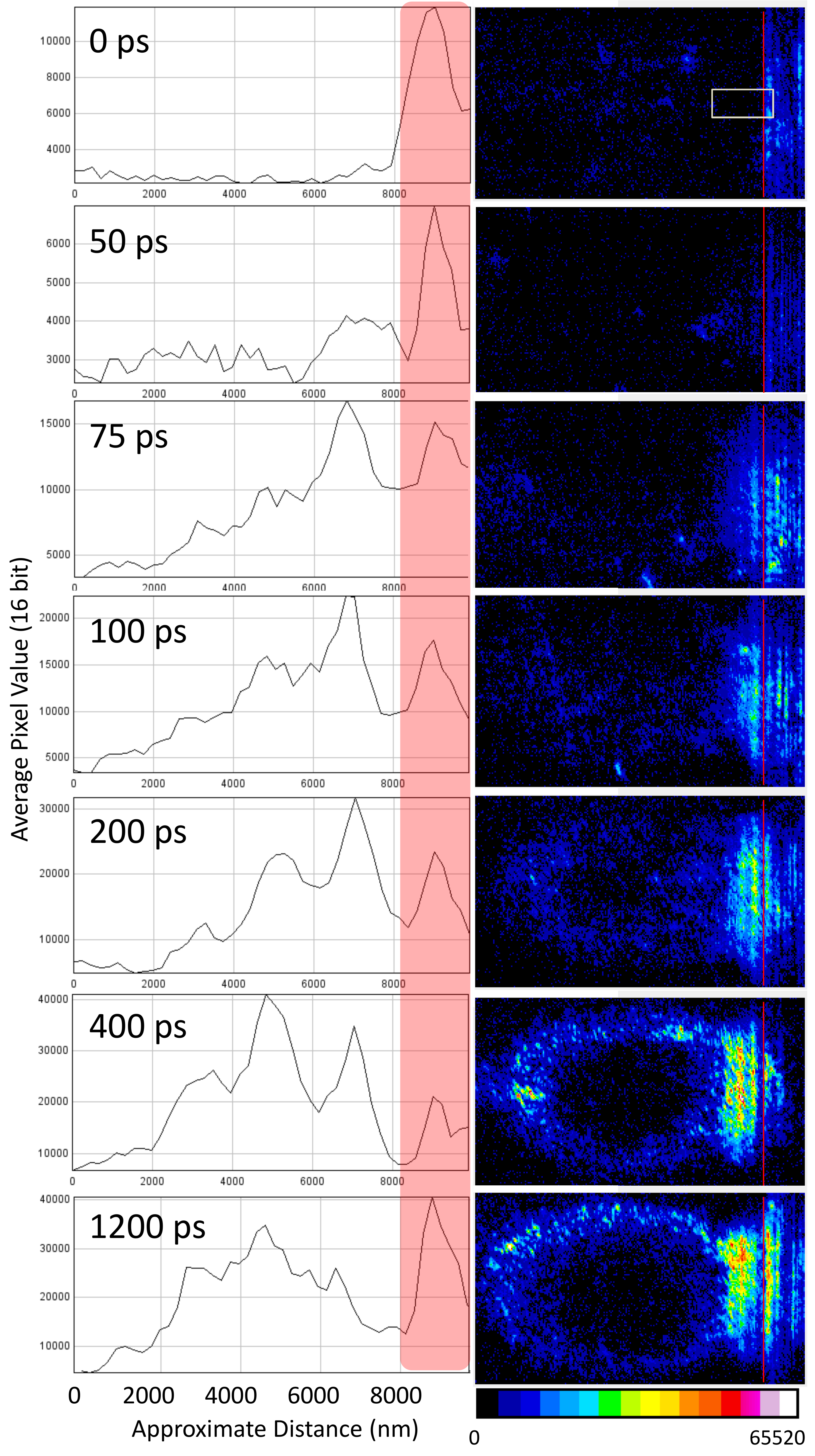}
\caption{\label{fig:trdm}Time-resolved diffraction microscopy images and corresponding lineouts. Images are in false color with pre-fabricated groove edge marked (red line to guide the eye), and averaged horizontal lineout location shown (gray rectangle). LIPSS are beginning to form near 50-75 ps, are well-developed by 200-400 ps, after which they are deformed by the central spot evolution. Horizontal axis is labeled approximate due to magnification uncertainty induced by oblique microscope objective.}
\end{figure}

Fig. \ref{fig:trdm} displays the data collected by this time-resolved diffraction microscopy method. Each false-color image was collected by initiating damage at a new site along the fabricated groove, with the probe set to the indicated delay time. The Gaussian maximum of the pump pulse can be seen by the elliptical feature, which is off-center from the groove by $\SI{19}{\micro\meter}$. The corresponding plots are lineouts of the pixel values along the central axis, and the red shading indicates the groove location. LIPSS are beginning to form at 50-75 ps, consistent with Ref. \cite{Murphy2013pumpprobe}, and are well developed by 200-400 ps. At longer delays, the LIPSS become more distorted, and by 1.200 ns the structure appears very similar to its final state (not shown). As far as we are aware, this is the first reported observation of LIPSS formation in which the formation of each period is resolved separately.
\begin{figure}
\includegraphics[scale=.275]{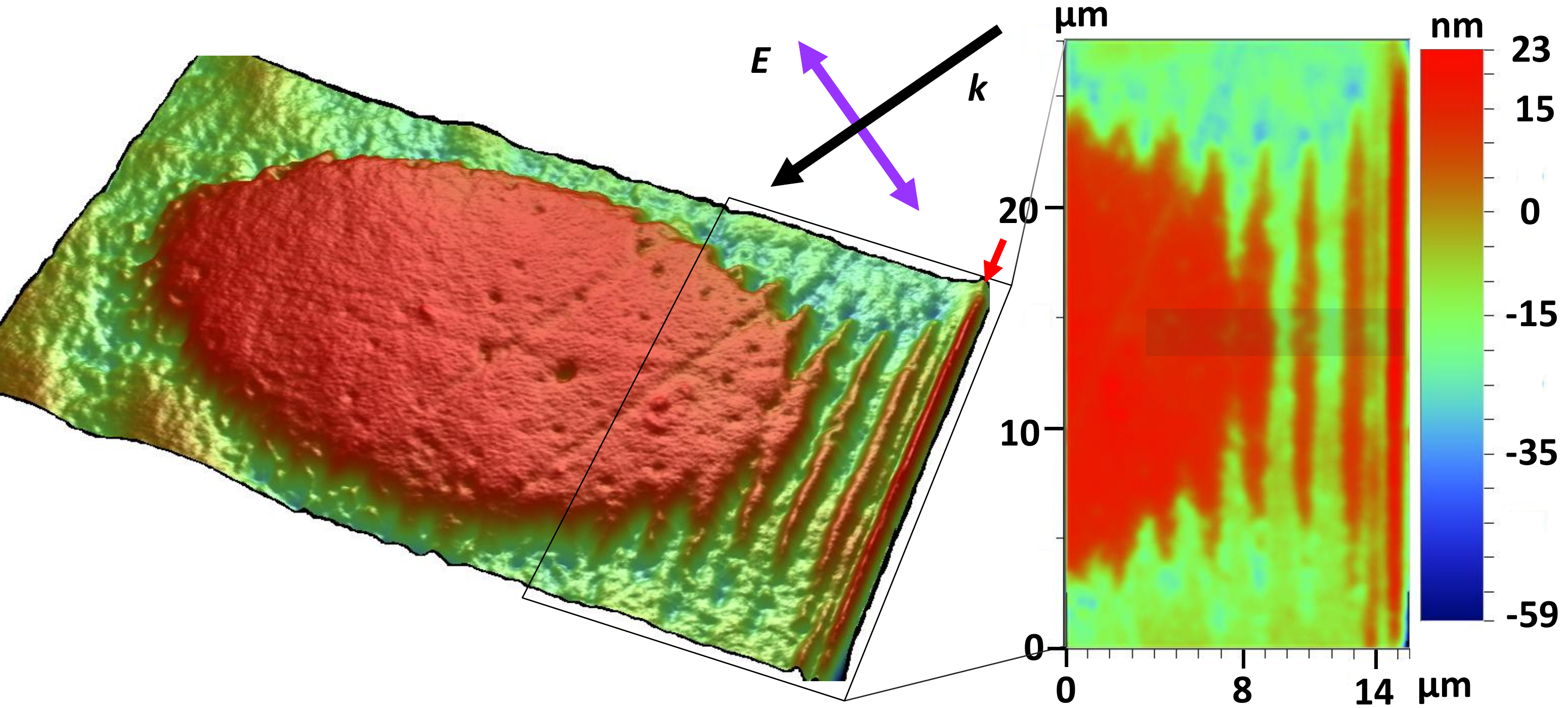}
\caption{\label{fig:wyko}Study of damage site resulting from femtosecond pulse interaction with grooved single crystal Cu substrate, profiled with interferometric microscope. (a) 3D profile rendered with depth dimension exaggerated by $\sim$300x. Incident laser $\boldsymbol{k}$ ($\SI{45}{\degree}$) and E (p-pol.) shown. Damaged area is raised (crater formation absent), and mostly flat, which is consistent with the dark central spots observed forming in TRDM images. (b) 2D false color depth profile, with faint gray bar indicating lineout shown in Fig. \ref{fig:heating}. LIPSS period $\SI{2.3}{\micro\meter}$. Peak at far right (red arrow) marks the location of the initial groove edge, but it clearly has also been modified by the laser.}
\end{figure}

We used an interferometric microscope (Wyko NT9100 by Veeco) to profile several damage sites, and an example trace is shown in Fig. \ref{fig:wyko}. The profile shows an elliptical central spot which is reasonably flat, in addition to the LIPSS which were seen in the TRDM images. This flatness would explain why the central region has low diffracted signal in TRDM. All of these features are raised above the initial sample elevation, which suggests that at this low fluence there is minimal ablation. Instead, we interpret that the copper is melted and amorphized in these regions, and due to an expected approximate $4\%$ volume increase for amorphous copper \citep{Brown1995amorph}. A $\SI{30}{\nano\meter}$ rise then implies about $\SI{700}{\nano\meter}$ melt depth, and with an elliptical melt area of about $\SI{700}{\micro\meter^2}$ we can estimate required energy to simultaneously melt the whole volume $U_{melt}\approx\SI{0.9}{\micro\joule}$, which is only $\sim1/5$ of the incident energy. LIPSS period is measured to be $\SI{2.3}{\micro\meter}$, which is consistent with the prevalent SPP model for this AOI \citep{Huang2009}. The LIPSS period is given by $\Lambda=\lambda/(\lambda/\lambda_s\pm \sin\theta)$ with AOI $\theta$, SPP wavelength $\lambda_s=\lambda\sqrt{(\epsilon'+\epsilon_d)/\epsilon'\epsilon_d}$, and $+$ or $-$ determining backward- or forward-propagating SPPs respectively.  In this case, $\epsilon_d=1$ for air and $\epsilon=\epsilon'+i\epsilon''=-22.7+i1.89$ the dielectric function for copper, which gives $\Lambda=2.45^{+0.20}_{-0.18} \SI{}{\micro\meter}$ due to uncertain $\theta=45\pm\SI{2}{\degree}$ from alignment.

The laser-induced damage process is generally understood as a fluence distribution which excites electrons in the interaction regime during the laser pulse, and then later transfers energy to the lattice via electron-ion collisions (i.e. the two-temperature model \cite{Jiang2005}). After the ions are heated, they can propagate heat via ion-ion collisions and/or undergo phase transitions which change the surface morphology. These transport processes have not yet been well characterized, but determining the cause of our time-ordered LIPSS formation result may give insight into the excited metal dielectric function $\epsilon_{hot}$, the Drude collision frequency $\Gamma$, and the electron temperature $T_e$.

\begin{figure}
\includegraphics[scale=.32]{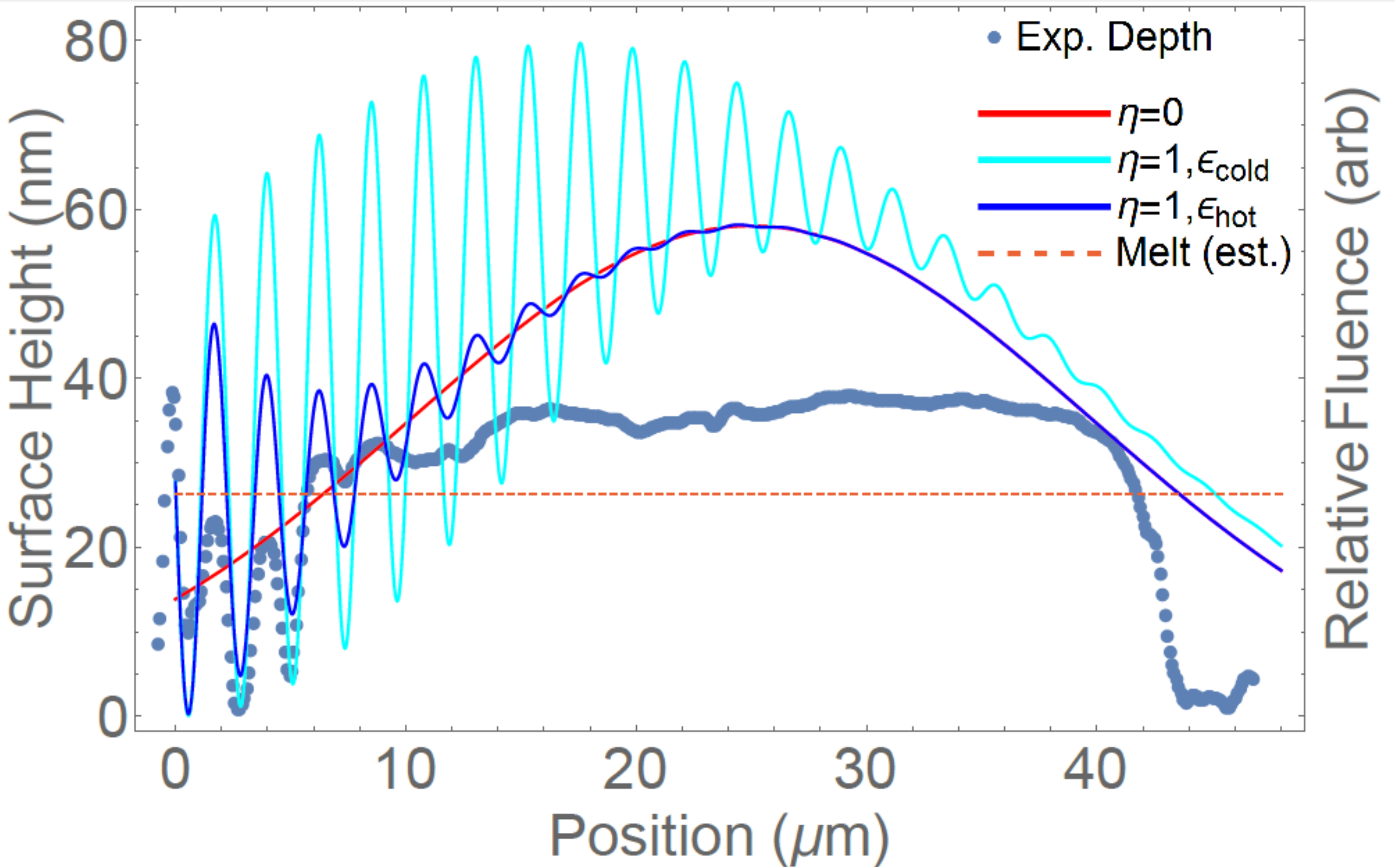}
\caption{\label{fig:heating}Results of Eq. \ref{eq:intE2} the 1D electron heating model (lines), and experimental depth profile (dots). The laser heating (red) in the absence of an SPP ($\eta=0$) is consistent with the central spot which forms with or without the surface scratch, and a relative melt threshold can be estimated (dashed). For SPP parameters determined by $\epsilon_{cold},\eta=1$ (cyan), neither the hot-groove temperature gradient, nor the laser spot profile are reproduced. The experimental results are better represented using $\epsilon_{hot}$ (blue), with temperature gradient toward the scratch, and the correct number of LIPSS are predicted before they are washed out by the central melt zone.}
\end{figure}

Our observations show a surprising result, that formation of LIPSS from a pre-existing groove is time-ordered such that the structures form sequentially away from the source groove, even though the laser intensity there is the lowest. To explain the time-ordering, we postulate that the LIPSS ridge nearest to the groove had the highest $T_e$, causing it to undergo ultrafast melting and expansion faster than subsequent LIPSS ridges. This distribution assumes $T_e$ gradient is opposite the to the laser's Gaussian intensity gradient, due to the SPP electric field enhancement. It also assumes that energy stored in the combined E-field excites electrons, which thermalize within themselves via electron-electron collisions, which, in turn transfer heat to local ions 'adiabatically'. Based on these assumptions, we construct a 1-D model where the near surface electron energy distribution in a direction perpendicular to the nano-groove is calculated by spatio-temporally integrating the square of the laser and the SPP E-field originating from the nano-groove (energy stored) over the laser pulse:
\begin{equation}\label{eq:intE2}
F(x)=\int_{-\infty}^\infty I(x) dt \propto \int_{-\infty}^\infty (\boldsymbol{E_{Laser}} +\eta\boldsymbol{E_{SPP}})^2 dt
\end{equation}
where $\eta$ is the relative field amplitude between the SPP and laser, and the normalized field terms are given by
\begin{align}
\boldsymbol{E_{Laser}}=&\exp[-(\frac{x-v_Lt}{v_L\tau})^2-(\frac{x-d}{w_0\sin\theta})^2 \nonumber\\ &+i(k\sin\theta x -i\omega t)] \\
\boldsymbol{E_{SPP}}=&\exp[-(\frac{x-v_st}{v_s\tau})^2-(\frac{-d}{w_0\sin\theta})^2 \nonumber \\ &+i(\operatorname{Re}[k_s] x -i\omega t)]-\operatorname{Im}[k_s]x] \label{eq:Espp}
\end{align}
where $v_L=c/\sin\theta $ and $v_s=c/\sqrt{(\epsilon'+\epsilon_d)/\epsilon'\epsilon_d}\approx0.98c$ are respectively the laser and SPP phase velocities along the surface, $\tau$ the pulsewidth, $d$ the laser off-center distance, and $w_0$ the beam waist radius. For both fields, the first term is the propagating temporal profile, the second term is the laser spatial profile which is simply an amplitude for the SPP, the third is the phase term, and the fourth term in Eq. \ref{eq:Espp} is the SPP decay due to propagation. Eq. \ref{eq:intE2} can be integrated analytically (not shown), and the results are shown in Fig. \ref{fig:heating} for our experimental parameters and various $\eta$ and $\epsilon$.

The field enhancement $\eta$ has been calculated previously for a plane wave \citep{Weber1981}:
\begin{equation}
\eta^2=\frac{E_{SPP}^2}{E_{Laser}^2}=T^{el}=\frac{2\left|\epsilon'\right|^2\cos\theta(1-R)}{\epsilon''(\left|\epsilon'\right|-1)^{1/2}}
\end{equation}
In our case this could give rise to values as high as $\eta\approx9(1-R)^{1/2}$ for room temperature copper, or $\eta\approx2(1-R)^{1/2}$ if heating were to increase $\Gamma$ by an order of magnitude so that $\epsilon_{hot}\approx-12+i12$, using simple equation for the dielectric function of a metal: $\epsilon_{metal}=1-\omega_p^2/{\omega(\omega+i\Gamma)}$ . Fig \ref{fig:heating} shows that the 1-D model qualitatively explains our experimental observations best with $\epsilon_{hot}$, whose primary effect is reduction of the propagation length of the SPP by an order of magnitude, while only varying $\Lambda$ by $1\sim2\verb+%+$. Without this reduction of SPP propagation length ($\epsilon_{cold}$), the decay of the oscillations is dominated by the decreasing temporal overlap between the SPP and laser, arising due to velocity mismatch at this AOI. 

To arrive at this $\epsilon_{hot}$ value, we first calculate $\Gamma$ from the known $\epsilon$ at room temperature, and then add in the electron-electron collisions $\Gamma_{ee}$ as the electron temperature $T_e$ sharply increases due to the laser. At room temperature, electron-phonon collisions $\Gamma_{ep}$ dominate, but since the ion temperature remains basically unchanged during an ultrashort laser pulse, models \cite{LeeMore1984}\cite{Fisher2001} predict only a weak dependence of $\Gamma_{ep}$ on $T_e$. Then from $\epsilon_{metal}$ and the $\SI{300}{\kelvin}$ values we obtain $1/\Gamma_{ep}=\SI{5.14}{\femto\second}$. Electron-electron collisions $\Gamma_{ee}$ are given by \cite{Fisher2001}
\begin{equation}
\Gamma_{ee}\approx\frac{E_F}{\hbar}\frac{k_BT_e}{E_F}^2 \text{   for   } k_BT_e<E_F
\end{equation}
with Planck constant $\hbar$, Fermi energy $E_F=\SI{7.0}{\electronvolt}$ for copper, and Boltzmann constant $k_B$. While negligible at $\SI{300}{\kelvin}$ ($1/\Gamma_{ee}=\SI{6.9}{\pico\second}$), $\Gamma_{ee}$ begins to dominate at temperatures on the scale of the electron energy in laser field, $U_p=e^2E_{Laser}^2/(4m_e\omega^2)=\SI{0.53}{\electronvolt}$. Taking into account the off-center distance and contribution of the SPP field gives an effective $U_{p-eff}=\SI{1}{\electronvolt}$, which is still lower than the $T_e$ distribution of the blue curve shown in Fig. \ref{fig:heating}, which corresponds to $T_e=\SI{3}{\electronvolt}$ and $\Gamma=\Gamma_{ee}+\Gamma_{ep}=1/0.5\SI{}{\femto\second}$ and best matches final combined LIPSS and laser melt zone profile. As ponderomotive scaling allows for electron KE up to 10 $U_p$ \citep{Yang1993}, contributions from inter-band transitions ($\SI{2.2}{\electronvolt}$\citep{Fox2001}) due to conduction-band electrons collisionally ionizing valence electrons, could significantly change the electron density, resulting in higher $T_e$, which, in turn would raise $T_{ion}$ beyond melt threshold \cite{Jiang2005}.

In this letter we have presented the results of time-resolved diffraction microscopy experiment, which for the first time resolved the temporal dependence of LIPSS formation by a femtosecond laser pulse interacting with a surface groove. The LIPSS were observed to form sequentially in time with increasing distance from the groove. While many aspects of LIPSS dynamics need to be studied further, such as the spatiotemporal-dependence of $\epsilon$, details of the transport process, our 1D model appears to  capture most of the relevant features: LIPSS period, temperature gradient of LIPSS toward the groove, central melt size, and even an order of magnitude estimate of $T_e$. This technique can be used widely across materials to understand LIPSS formation dynamics in general, and evolution of di-electric response of materials in particular. Overall, our model of electron heating due to local field energy density provides a reasonable explanation to the experimental observations. This work also motivates the need of further theoretical and computational efforts to calculate dielectric function of solids under intense fields.  

This work was supported by the Air Force Office of Scientific Research grant no. FA9550-12-1-0454.

\bibliography{bibfile}

\end{document}